\documentclass[11pt,twoside]{article}
\usepackage{./asp2014}

\aspSuppressVolSlug
\resetcounters

\begin{document}

\title{Central stars of mid-infrared nebulae discovered with {\it Spitzer} and {\it WISE}}
\author{V.V.~Gvaramadze$^{1,2}$ and A.Y.~Kniazev$^{1,3,4}$
\affil{$^1$Sternberg Astronomical Institute, Lomonosov Moscow
State University, Universitetskij Pr.~13, Moscow 119992, Russia;
\email{vgvaram@mx.iki.rssi.ru}} \affil{$^2$Space Research
Institute, Russian Academy of Sciences, Profsoyuznaya 84/32,
117997 Moscow, Russia} \affil{$^3$South African Astronomical
Observatory, PO Box 9, 7935 Observatory, Cape Town, South Africa;
\email{akniazev@saao.ac.za}} \affil{$^4$Southern African Large
Telescope Foundation, PO Box 9, 7935 Observatory, Cape Town, South
Africa}}

\paperauthor{V.V.~Gvaramadze}{vgvaram@mx.iki.rssi.ru}{}{Lomonosov Moscow
State University}{Sternberg Astronomical Institute}{Moscow}{}{119992}{Russia}
\paperauthor{A.Y.~Kniazev}{akniazev@saao.ac.za}{}{South African Astronomical
Observatory}{}{Cape Town}{}{7935}{South Africa}

\begin{abstract}
Searches for compact mid-IR nebulae with the {\it Spitzer Space Telescope} 
and the {\it Wide-field Infrared Survey Explorer} ({\it WISE}), accompanied 
by spectroscopic observations of central stars of these nebulae led to the 
discovery of many dozens of massive stars at different evolutionary stages, 
of which the most numerous are candidate luminous blue variables (LBVs). In 
this paper, we give a census of candidate and confirmed Galactic LBVs revealed 
with {\it Spitzer} and {\it WISE}, and present some new results of 
spectroscopic observations of central stars of mid-IR nebulae.
\end{abstract}

\section{Introduction}\label{int}
The majority of massive stars reside in binary systems \citep{Sa12}
and can lose a significant fraction of their initial
mass because of interaction with their companion stars through
mass transfer, common envelope evolution or merger. In addition to
binary interaction processes, the copious stellar winds of the
most massive stars also significantly reduce their initial masses. 
The material lost by massive stars can, in principle, be detected 
as parsec-scale circumstellar nebulae. But since these stars form in 
star clusters and the majority of them end their lives there, the 
origin of coherent (observable) nebulae around stars residing in their 
parent clusters is hampered by the effect of stellar winds and ionizing
emission from nearby massive stars. On the other hand, about 20\%
of OB stars escape from their birth places because of dynamical
encounters with other massive stars or because of supernova explosions 
in binary systems \citep{Bl61,Gi87}. These so-called runaway stars are 
more suitable for producing observable circumstellar nebulae because 
they are not exposed to destructive influence of nearby massive stars.
Indeed, the vast majority of known circumstellar nebulae are
associated with stars located either in the field (e.g. nebulae 
around the bona fide LBVs AG\,Car and HR\,Car) or at the
outskirts of their parent clusters (e.g. the Pistol Nebula).

Among stars of different evolutionary stages the circumstellar
nebulae are very common for LBVs, of which more than 70\% have 
associated nebulae \citep*{Kn15b}. The 
circumstellar nebulae can also be found around late nitrogen sequence 
Wolf-Rayet stars and blue supergiants (BSGs), and less frequently around 
red supergiants, Be and B[e] stars. The detection of nebulae reminiscent 
of those associated with known massive stars serves as a strong indication 
that their central stars are massive as well. Indeed, follow-up spectroscopy 
of stars selected in this way has led to the discovery of many dozens of 
massive stars at various evolutionary stages \citep{Cl03,Gv09,Gv10a,Gv10b,Gv12,
Gv14a,Gv14b,Gv15a,Gv10c,Gv15b,Wa10,Wa11,Ma10,St12a,St12b,Bu13,Fl14,Kn15a,Kn15b,Kn16}. 
Given the high percentage of LBVs having circumstellar 
nebulae, it is not surprising that many of the newly identified massive stars
show LBV-like spectra and therefore were classified as candidate LBVs (cLBVs). 
In this paper, we give a census of (c)LBVs revealed with {\it Spitzer} and
{\it WISE}, and present some new results of spectroscopic observations of 
central stars of mid-infrared (mid-IR) nebulae.

\section{Mid-IR nebulae discovered with {\it Spitzer} and {\it WISE}}
\label{neb}
Because of high visual extinction in the Galactic plane -- close
to which the majority of massive stars are concentrated -- the
most effective way for detection of circumstellar nebulae is through 
mid-IR observations. Using data from various {\it Spitzer} Legacy
Programmes\footnote{http://sha.ipac.caltech.edu/applications/Spitzer/SHA/}
and the {\it WISE} all-sky survey \citep{Wr10}, we
detected about 250 nebulae with central stars. These nebulae have
a wide range of morphologies, ranging from circular and ring-like
to bipolar and more complex forms (see Fig.\,\ref{lbv}). 115 of them 
were discovered with {\it Spitzer} by 2010 and presented in \citet*{Gv10c},
while others were detected later on with either {\it
Spitzer} or {\it WISE}. Almost all of the nebulae were found in
the {\it Spitzer} 24\,$\mu$m or {\it WISE} 22\,$\mu$m images. A two 
times lower angular resolution of the {\it WISE} data resulted in that
the nebulae detected with this telescope have, in general, larger
angular dimensions, which suggests that they are located at closer
distances. This is supported by the fact that the percentage of
{\it WISE} nebulae having optical counterparts is higher than that
among the nebulae detected with {\it Spitzer} (see Fig.\,\ref{neb} for
three examples of such {\it WISE} nebulae).

\articlefigure[width=0.75\textwidth]{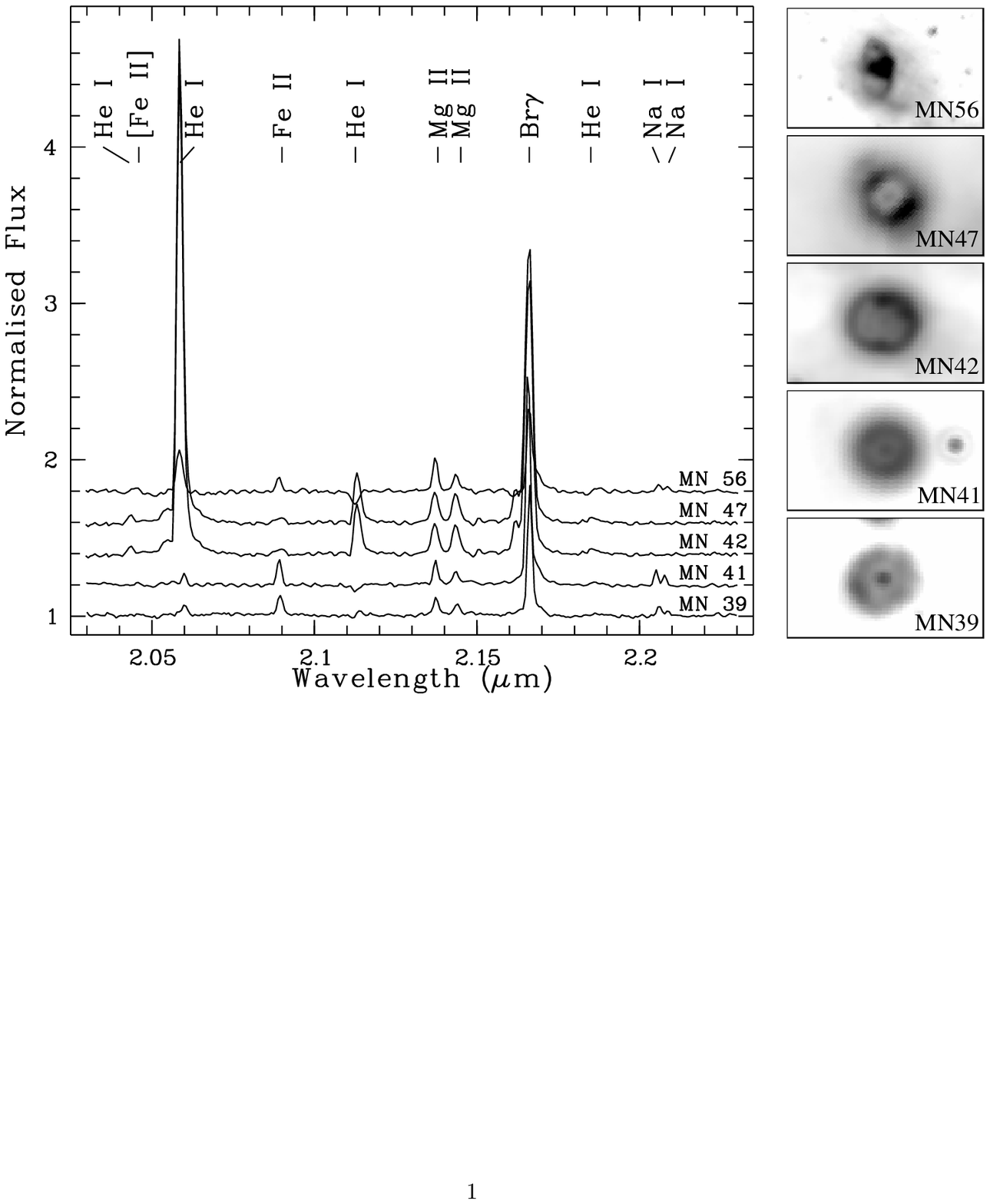}{vlt}{VLT/ISAAC $K$-band spectra 
(left panel) of cLBVs associated with mid-IR nebulae (right panel).}

\section{Central stars of mid-IR nebulae discovered with {\it Spitzer} and {\it WISE}}
\label{neb}
The majority of the central stars are either invisible or dim in the optical
band. To determine their nature, one needs therefore to use IR telescopes or 
large (8--10-m class) optical ones. Fig.\,\ref{vlt} (right panel) 
shows {\it Spitzer} 24\,$\mu$m images of five nebulae, whose central
stars are visible only in the IR and whose $K$-band spectra, taken
with the ISAAC spectrograph on the ESO Very Large Telescope 
(VLT), are presented in the same figure (left panel). Interestingly, although 
all five nebulae look quite different, their spectra are very similar
to each other and to those of known LBVs \citep[see e.g. fig.\,2 in][]{St12b}, 
which suggests a classification of these stars as cLBVs.

\begin{table}
  \caption{Candidate and bona fide LBVs discovered with {\it Spitzer} and {\it WISE}.}
  \label{tab:cLBV}
  \renewcommand{\footnoterule}{}
  \begin{center}
  \begin{minipage}{\textwidth}
  \begin{tabular}{llllll}
      \hline 
{\it cLBVs}: & & & & & \\
MN1$^a$ & Hen\,3-729$^b$ & MN7$^b$ & MN8$^a$ & MN13$^c$ & MN30$^c$ \\
MN39$^c$ & IRAS\,16115$-$5044$^b$ & MN41$^c$ & WMD14$^a$ & MN42$^{c,d}$ & MN45$^a$ \\
MN46$^e$ & MN47$^c$ & MN51$^a$ & MN53$^c$ & MN55$^a$ & MN56$^f$ \\
WS2$^g$ & MN58$^{c,h}$ & MN61$^{c,h}$ & MN64$^{a}$ & MN68$^{a,c}$ & MN76$^d$ \\
MN79$^{c,i}$ & MN80$^h$ & MN81$^i$ & MN83$^h$ & MN84$^{a,h}$ & MN87$^j$ \\
MN90$^c$ & IRAS\,18433$-$0228$^i$ & MN96$^{c,d}$ & MN101$^{c,d}$ & MN107$^i$ & MN112$^k$ \\
{\it LBVs}: & & & & & \\
WS1$^l$ & Wray\,16-137$^m$ & MN44$^n$ & MN48$^o$ & & \\
\hline
    \end{tabular}
    \end{minipage}
    \end{center}
     $^a$\citet{Wa10}; $^b$\citet{Kn15a}; $^c$\citet{Wa11}; $^d$\citet{St12a}; 
     $^e$\citet*{Gv10c}; $^f$\citet{Gv15a}; $^g$\citet{Gv12}; $^h$\citet{St12b}; 
     $^i$\citet{Fl14}; $^j$\citet{In16}; $^k$\citet{Gv10b}; $^l$\citet{Kn15b};
     $^m$\citet{Gv14b}; $^n$\citet*{Gv15b}; $^o$\citet{Kn16}.    
    \end{table}

Optical and IR spectroscopy of other central stars, carried out by our and two 
other independent groups, led to the discovery of 40 cLBVs, which doubled the 
number of known Galactic stars of this type \citep[cf.][]{Cl05}. 
Moreover, follow-up spectroscopic and photometric monitoring of some of these cLBVs 
revealed significant changes in the spectra and brightness of four of them, 
Wray\,16-137 \citep{Gv14b}, WS1 \citep{Kn15b}, MN44 \citep*{Gv15b} and MN48 \citep{Kn16}, 
implying that these stars belong to the family of Galactic bona fide LBVs, which 
currently consists of 18 members \citep[see table\,2 of][]{Kn15a}. 
In Table\,1 we give a current census of candidate and bona fide LBVs discovered 
with {\it Spitzer} and {\it WISE}. 

In Fig.\,\ref{lbv} (left panel) we show images of nebulae around 36 cLBVs to
further illustrate the variety of shapes inherent to these nebulae. In the 
same figure (right panel), we also show nebulae around 6 BSGs: from up to 
bottom, respectively, WS3 (or BD+60$\deg$\,2668), MN18, WS29 (or ALS\,19653), 
MN94, MN108 and MN113. One can see that the nebulae associated with both types 
of stars show a wide range of morphologies and that individual objects from one 
group of nebulae have close counterparts among nebulae from the other group. 
Possible implications of this similarity is that (i) some BSGs might be 
dormant LBVs, (ii) nebulae around some (c)LBVs might be formed during the 
preceding evolutionary stages and therefore their origin is not necessarily 
related to episodes of enhanced mass loss experienced by LBVs. Also, the 
variety of shapes inherent to nebulae around (c)LBVs and BSGs suggests that 
several mechanisms might be responsible for their formation and, perhaps,
for triggering the LBV activity of massive stars. 

Finally, in Figs\,\ref{bsg-spec} and \ref{neb} we present spectra and images at 
several wavelengths of two of the six BSGs shown in Fig.\,\ref{lbv}. Both these BSGs,
WS3 and WS29, were discovered through detection of their circumstellar nebulae 
with {\it WISE} and both nebulae have optical counterparts, detected respectively
in the INT Photometric H$\alpha$ Survey of the Northern Galactic Plane 
\citep[IPHAS;][]{Go08} and the Digitized Sky Survey\,II \citep[DSS-II;][]{Mc00}. 
Note that the central star of WS3 consists of two components 
(separated by about 3 arcsec), which we classify as B0\,Ib and B5-6\,Ib. Note 
also that WS29 is indicated in the SIMBAD data base as a planetary nebula (PN), 
while our spectroscopy shows that it is a B0\,Ib star.

\articlefigure[width=0.75\textwidth]{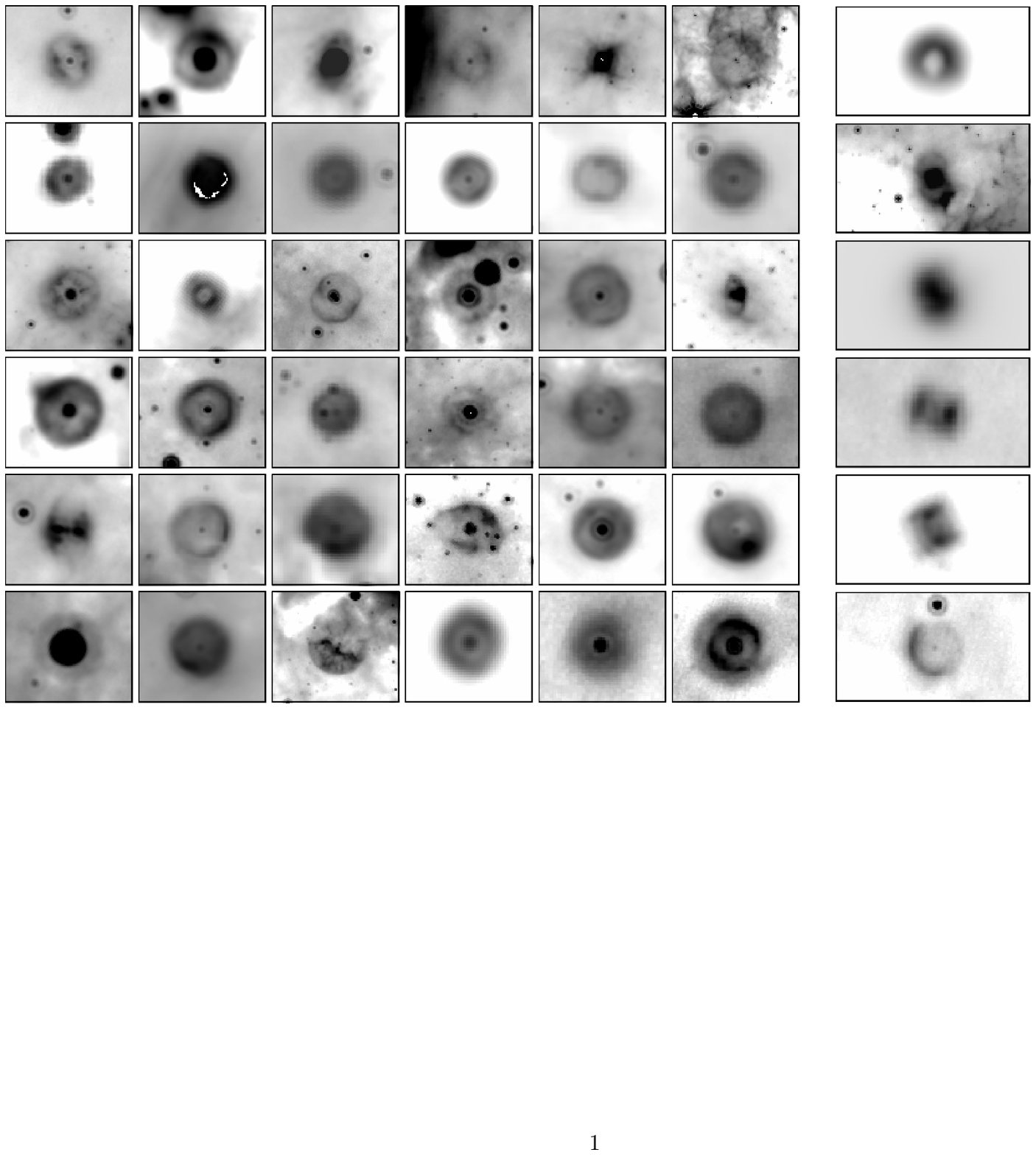}{lbv}{{\it Spitzer} and {\it WISE} 
images of nebulae associated with 36 cLBVs (shown in the left panel in the same 
order as they appear in Table\,1) and 6 BSGs (right panel). See text for details.}

\articlefigure[width=0.6\textwidth,angle=270,clip=0]{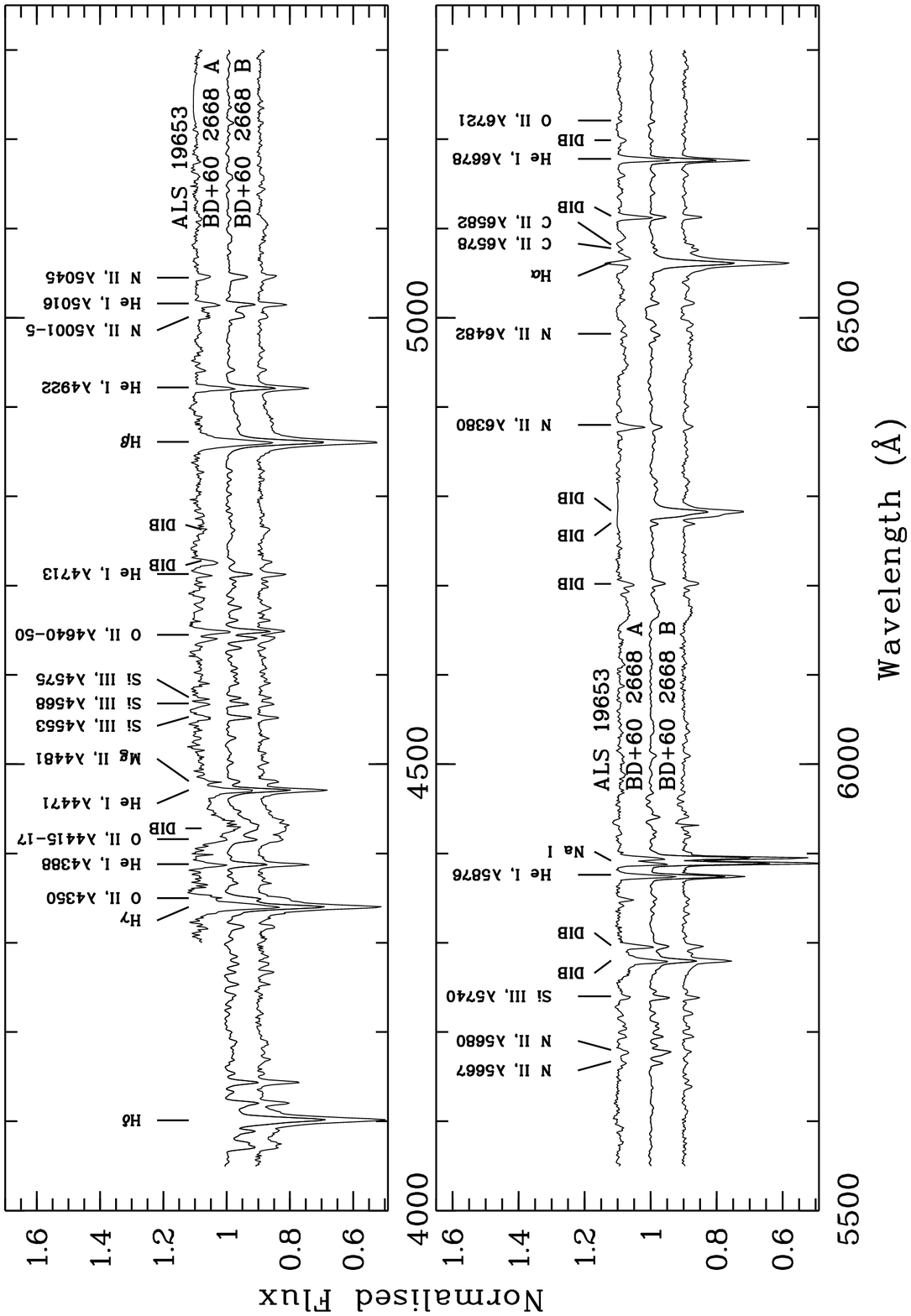}{bsg-spec}{Normalized 
spectra of WS29 (ALS\,19653) and two components of WS3 (BD+60$\deg$\,2668) obtained
respectively with the Southern African Large Telescope and the 3.5-m telescope in 
the Observatory of Calar Alto (Spain).}

\articlefigurethree{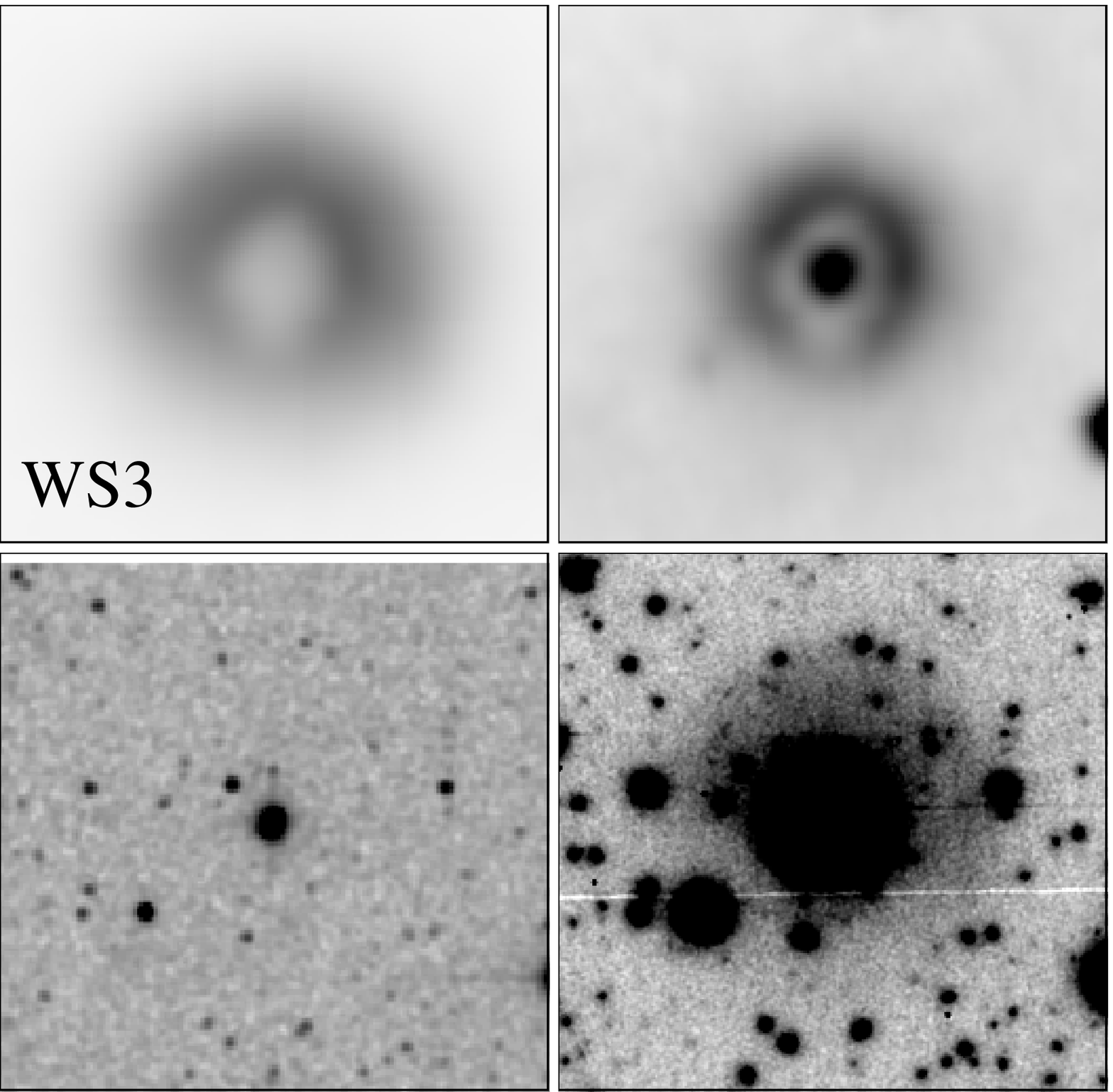}{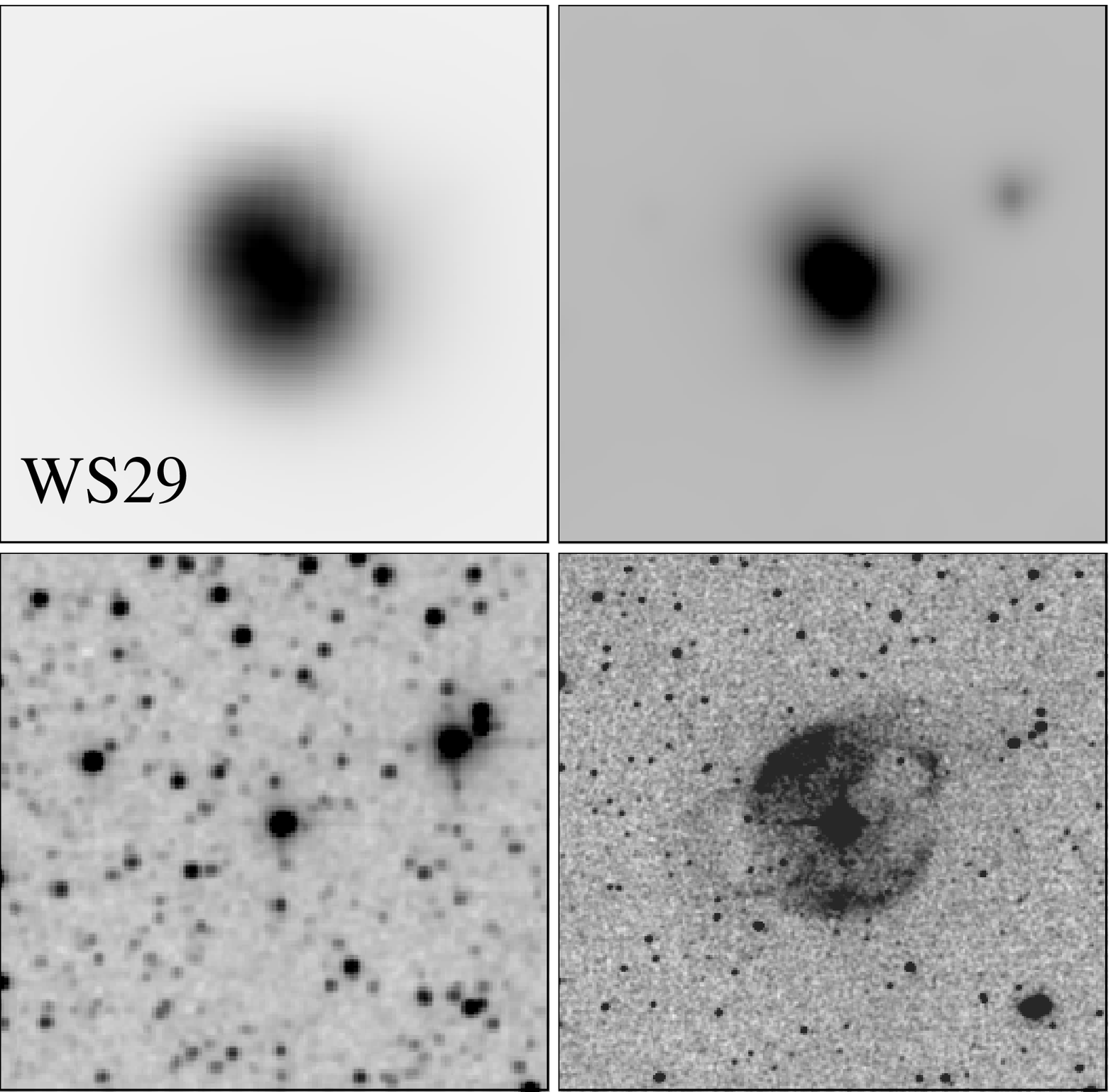}{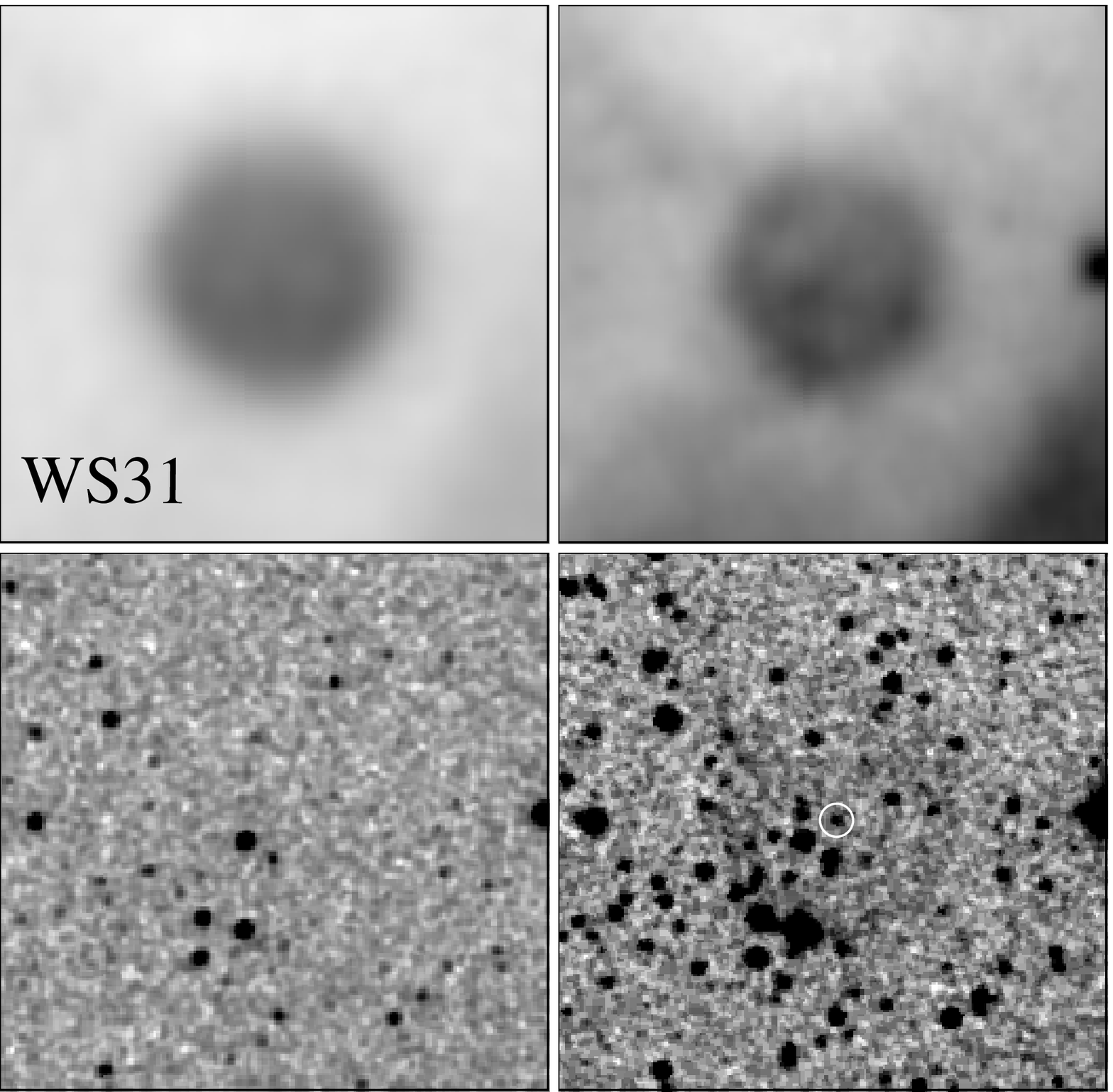}{neb}{Multi-wavelength images of three
nebulae discovered with {\it WISE}. Upper rows show {\it WISE} 22 (left) and 12\,$\mu$m 
images of each nebula. Left panels in the bottom rows are the 2MASS K-band images, while 
the right panels are the IPHAS H$\alpha$ (WS3 and WS29) and the DSS-II red band (WS31)
images. See text for details.}

\section{By-products}\label{pn}
It should be noted that only two of the newly-discovered mid-IR nebulae with 
central stars, namely MN102 and WS31, turn out to be PNe, and both of them have 
optical counterparts. MN102 was discovered with {\it Spitzer} and its central 
star is of [WC] type \citep[Gvaramadze et al. 2010c; see also][]{Fl14}.  
WS31 was detected with {\it WISE}. In the 22 and 12\,$\mu$m 
images it appears as a circular nebula without a central star 
(Fig.\,\ref{neb}, right panel), which is also not visible in other {\it WISE} and
all 2MASS \citep[Two Micron All-Sky Survey;][]{Sk06} wavebands. 
Surprisingly, we found that this nebula has an 
obvious, previously unknown, counterpart in the DSS-II red band image. 
We identified the central star of the nebula 
with a blue star at RA(J2000)=$22^{\rm h} 27^{\rm m} 39\fs19$ and 
Dec.(J2000)=$+66\deg 44\arcmin 09\farcs6$ (marked in the DSS-II image 
by a white circle). The Seventh Data Release of the Sloan Digital Sky 
Survey \citep{Ab09} provides for this star
the following photometry: $u$=21.116$\pm$0.115 mag, $g$=20.540$\pm$0.027 mag,
$r$=20.087$\pm$0.025 mag, $i$=19.846$\pm$0.028 mag and $z$=19.706$\pm$0.090 mag.
To our knowledge, this is the first new PN detected with {\it WISE}.

\section{Conclusions}

\noindent
$\bullet$ Mid-IR imaging provides a powerful tool for revealing LBVs 
and related stars.\\
$\bullet$ Spectroscopic and photometric observations of the newly-identified
cLBVs continue, with discoveries of new bona fide LBVs anticipated.
\vspace*{0.5cm}

\acknowledgements This work is supported by the Russian Foundation
for Basic Research grant 16-02-00148. AYK also acknowledges
support from the National Research Foundation (NRF) of South
Africa.

\end{document}